\input{aipcheck}

\documentclass[
    ,final            
    ,numberedheadings 
    ,sort&compress
  ]
  {aipproc}

\usepackage{amssymb}
\usepackage{amsmath}

\layoutstyle{8x11single}

\begin{document}

\title{UHE neutrino and cosmic ray emission from GRBs:\\ revising the models and clarifying\\ the cosmic ray-neutrino connection}

\classification{95.85.Ry,98.70.Sa,98.70.Rz}
\keywords      {neutrino, cosmic ray, UHE, GRB}

\author{Mauricio Bustamante$^\star$}{
  address={Institut f\"ur Physik und Astrophysik, Universit\"at W\"urzburg, 97074 W\"urzburg, Germany}
}

\author{Philipp Baerwald}{
  address={Institut f\"ur Physik und Astrophysik, Universit\"at W\"urzburg, 97074 W\"urzburg, Germany}
  ,altaddress={Department of Astronomy and Astrophysics; Department of Physics; Center for Particle and Gravitational Astrophysics; Institute for Gravitation and the Cosmos; Pennsylvania State University, 523 Davey Lab, University Park, PA 16802, USA} 
}

\author{Walter Winter}{
  address={Institut f\"ur Physik und Astrophysik, Universit\"at W\"urzburg, 97074 W\"urzburg, Germany}
}

\begin{abstract}
 Gamma-ray bursts (GRBs) have long been held as one of the most promising sources of
ultra-high energy (UHE) neutrinos. The internal shock model of GRB emission posits
the joint production of UHE cosmic rays (UHECRs, above $10^8$ GeV), photons, and
neutrinos, through photohadronic interactions between source photons and
magnetically-confined energetic protons, that occur when relativistically-expanding
matter shells loaded with baryons collide with one another. While neutrino
observations by IceCube have now ruled out the simplest version of the internal shock
model, we show that a revised calculation of the emission, together
with the consideration of the full photohadronic cross section and other particle
physics effects, results in a prediction of the prompt GRB neutrino flux that still
lies one order of magnitude below the current upper bounds, as recently exemplified
by the results from ANTARES. In addition, we show that by allowing protons to
directly escape their magnetic confinement without interacting at the source, we are
able to partially decouple the cosmic ray and prompt neutrino emission, which grants
the freedom to fit the UHECR observations while respecting the neutrino upper bounds.
Finally, we briefly present advances towards pinning down the precise relation
between UHECRs and UHE neutrinos, including the
baryonic loading required to fit UHECR observations, and we will assess the role that very large volume neutrino
telescopes play in this.\\\\
$^\star$Presenter. E-mail: mbustamante@physik.uni-wuerzburg.de
\end{abstract}

\maketitle


\section{Introduction}

For a hundred years now we have been measuring the properties of cosmic rays, particles of extraterrestrial origin than span twelve orders of magnitude in energy. While the low-energy end of the spectrum is believed to be created by local sources such as the Sun, the production mechanism and origin sites of the ultra-high-energy cosmic rays (UHECRs), with $10^8$ GeV and higher, remain unknown. On the other hand, Earth- and satellite-based instruments such as Swift and Fermi have been for several years consistently observing sources of X-ray and gamma-ray emission, in the GeV-TeV range. These include active galactic nuclei (AGN), with luminosities in excess of $10^{40}$ erg s$^{-1}$ and occasionally flaring activity, and gamma-ray bursts (GRBs), with typical luminosities of $10^{52}$ erg s$^{-1}$. Unlike AGN, GRBs are transient sources, and the vast majority of their luminosity is usually output during the first $\lesssim 10$ s of activity, making them the most luminous objects in the Universe. It is hence natural to consider them to potentially be the common production sites of UHECRs and of gamma-rays. As we will see, the smoking gun signature of such a connection would be the detection of UHE neutrinos produced as a result of the interaction of the cosmic-ray protons or nuclei among themselves or with the photons at the source. Note that in our computations, we have assumed UHECRs to be composed solely of protons; even though there are hints of a heavier composition at the highest energies from the Pierre Auger Observatory \cite{Abraham:2010yv}, these are still inconclusive.


\section{An updated neutron model of UHE neutrino production}

Models that predict the neutrino flux from GRBs exist since the late 1990s \cite{Waxman:1995dg,Waxman:1998yy}. The fireball model of GRBs postulates that a central compact object ejects lumps, or shells, of highly-relativistic, beamed plasma loaded with electrons, photons, and protons. After an initial acceleration phase, the shells reach constant velocities and collide among each other, in what is known as the internal shock model. At this point, protons have been shock-accelerated to energies of the order of $\lesssim 10^{12}$ GeV. Within the plasma lumps, very intense magnetic fields, of order $10^5$ G, prevent the protons from escaping. When the ejecta collide with each other, $p\gamma$ interactions take place, and UHE photons, neutrons, and neutrinos are created in the same set of processes. To a first approximation, the main contribution comes from the $\Delta^+\left(1232\right)$ resonance, {\it i.e.}, $p\gamma \to \Delta^+\left(1232\right) \to n\pi^+$ or $p\pi^0$.
The charged pions decay and create neutrinos through $\pi^+ \to \mu^+ \nu_\mu \to \bar{\nu}_\mu e^+ \nu_e \nu_\mu$, while the neutral pions create gamma-rays through $\pi^0 \to \gamma \gamma$. The neutron is no longer magnetically confined, and so escapes the plasma and later beta-decays via $n \to p e^- \bar{\nu}_e$; the secondary UHE proton propagates and is detected at Earth as a cosmic ray. After flavor mixing, equal number of neutrinos of each flavor is expected at Earth. Therefore, this ``neutron model'' of UHECR emission model, where cosmic ray production occurs by neutron escape only, predicts the detection of one $\nu_\mu$ per cosmic ray. Lately, however, this paradigm has been challenged \cite{Ahlers:2011jj}, since it conflicts both with the neutrino and the gamma-ray bounds. 

The IceCube collaboration recently published updated upper bounds on the stacked neutrino flux from a selection of 117 GRBs \cite{Abbasi:2012zw}. In this analysis, the neutrino flux expected from each GRB was calculated using the analytical prescription by Guetta {\it et al.} \cite{Guetta:2003wi}, in which the neutrino spectrum is normalized to the observed photon fluence of the GRB, {\it i.e.},
$\int_0^\infty dE_\nu ~E_\nu F_\nu\left(E_\nu\right) \propto \int_\text{1 keV}^\text{10 MeV} d\varepsilon_\gamma~ \varepsilon_\gamma F_\gamma\left(\varepsilon_\gamma\right) \equiv F_\gamma$.
The neutrino fluxes for all of the selected GRBs were computed and added up. By assuming that 667 observable bursts occur per year, a quasi-diffuse flux was extrapolated (see Fig.~1 in \cite{Abbasi:2012zw}) which exceeds the IceCube upper bound, a result that led the collaboration to conclude that the basic neutron model of GRBs is disfavored.


However, a more detailed, numerical computation results in a quasi-diffuse neutrino flux that is instead an order of magnitude below the IceCube upper bound, as shown in Fig.~3 of Ref.~\cite{Hummer:2011ms}. This recalculation, dubbed NeuCosmA (Neutrinos from Cosmic Accelerators), differs from the analytical calculation of Ref.~\cite{Guetta:2003wi} in two respects. Firstly, instead of normalizing the neutrino flux directly to the observed photon fluence, it is instead the power-law proton density at the source which is normalized, {\it i.e.}, $\int_0^\infty dE_p^\prime ~E_p^\prime N_p^\prime\left(E_p^\prime\right) \propto F_\gamma/f_e$,
where $f_e^{-1}$ is the ``baryonic loading'', that is, the ratio of energy in protons to energy in electrons, commonly assumed to have a value around $10$. On the other hand, the photon density $N_\gamma^\prime$ at the source is assumed to follow a broken power law, normalized to the observed gamma-ray fluence. At every stage in the calculation, redshift and Lorentz corrections are properly taken into account. After normalizing the proton and photon densities, photohadronic interactions take place at the source: secondaries such as the $\pi^+$ lose energy adiabatically and through synchrotron radiation, and are allowed to decay. The resulting injected neutrino flux thus depends on the product of the proton and photon densities. The second important difference relative to the analytical prescription is that more neutrino-production channels are included in addition to the $\Delta^+$ resonance, namely, higher resonances, multi-pion, kaon, and direct ($t$-channel) production. Additionally, the $p\gamma$ interactions are no longer assumed to occur only at the representative photon break energy, but rather the interaction takes place along the whole photon spectrum. The maximum proton energy $E_{p,\text{max}}$ is determined by a competition of the relevant energy loss processes: adiabatic, synchrotron, and photohadronic energy losses. Details of the calculation can be found in Refs.~\cite{Hummer:2010vx,Baerwald:2010fk,Baerwald:2011ee,Hummer:2011ms} (see also Ref.~\cite{He:2012tq}). Recently, NeuCosmA was used by the ANTARES collaboration to perform a search for a GRB neutrino signal \cite{Adrian-Martinez:2013sga}.

\section{Beyond the neutron model}

\begin{figure}
  \includegraphics[width=\textwidth]{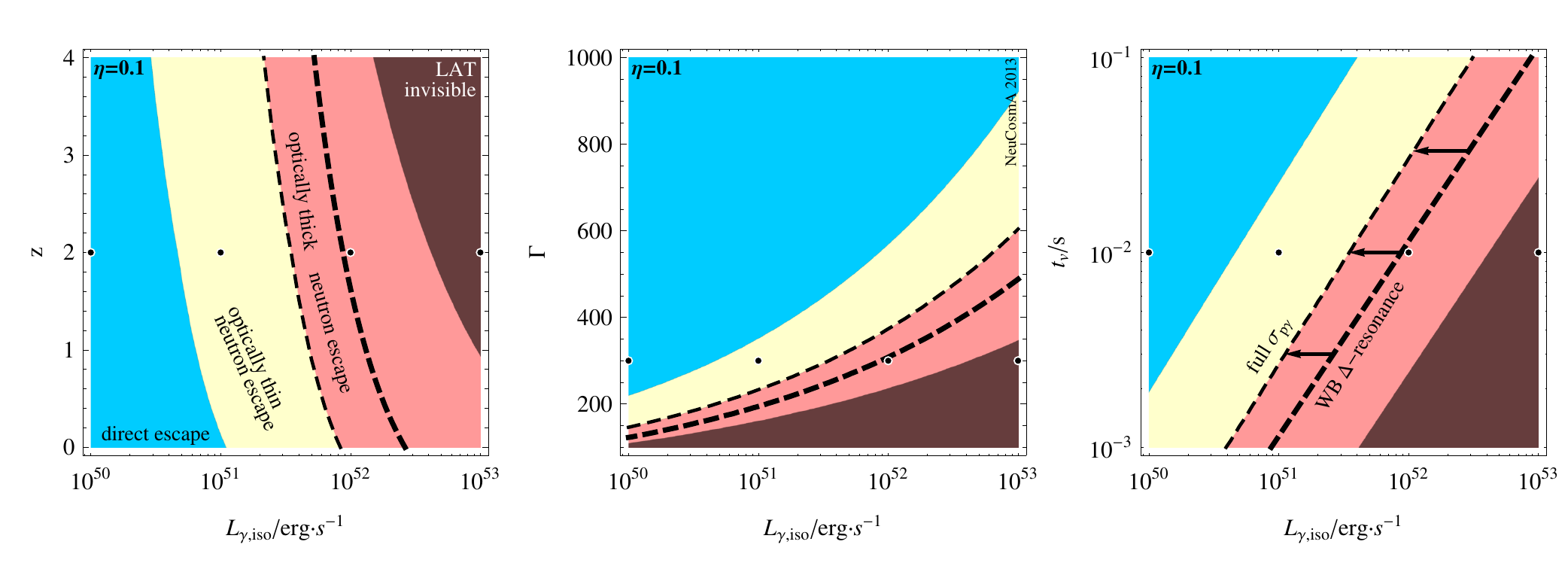}
  \caption{\label{fig:3regimeLisoscangrid2}Scans of the parameter space of UHECR emission from GRBs. Three regimes are identified: optically thin to neutron escape, {\it i.e.}, the neutron model (yellow); optically thick to neutron escape (red); and direct proton escape (blue). The brown region labeled ``LAT invisible'' represents the case where gamma-rays above $30$ MeV cannot leave the source due to pair production. Figure taken from Ref.~\cite{Baerwald:2013pu}.}
\end{figure}

The neutron model hinges on two basic assumptions: (i) protons are magnetically confined in the source, and only neutrons escape, and (2) protons interact at most once, while neutrons do not interact, {\it i.e.}, sources are optically thin to neutron escape. In Ref.~\cite{Baerwald:2013pu} we explored a further generalization of the neutrino prediction from GRBs that involves relaxing these two restrictions: on the one hand, if protons are not perfectly confined in the source, they will leak out without interacting and creating neutrinos, while, on the other hand, if protons interact multiple times in the source, the neutrino flux will be enhanced. Thus the ``one neutrino per cosmic ray'' paradigm of the neutron model is overcome. The optical depth to neutron escape is calculated as $\tau_n = \left. \left( t_{p\gamma}^{-1} / t_\text{dyn}^{-1} \right) \right\vert_{E_{p,\text{max}}}$, where $t_{p\gamma}$ is the photohadronic interaction timescale and $t_\text{dyn}$ is the time that the proton would take to free-stream through the plasma. Optically thin sources are those in which $\tau_n \lesssim 1$, while optically thick sources are those with $\tau_n > 1$. Furthermore, we allow for imperfect proton confinement: the protons that lie at a distance equal to one Larmor radius or less from the shell surface will leak out of it before interacting, thus leading to a decrease in the neutrino emission.

The value of $\tau_n$ and the fraction of directly-escaping protons depend on GRB parameters such as the observed luminosity $L_{\gamma,\text{iso}}$, redshift $z$, Lorentz factor of the expanding plasma $\Gamma$, and the variability timescale of the observed GRB lightcurve $t_v$. As shown in Fig.~\ref{fig:3regimeLisoscangrid2}, at different points in parameter space the GRB will be either optically thin to neutron escape (the neutron model), optically thick to neutron escape, or dominated by direct proton escape. Note that Fig.~\ref{fig:3regimeLisoscangrid2} has been generated for a low acceleration efficiency of $\eta = 0.1$ (defined via the acceleration timescale $t_\text{acc}^\prime = E^\prime/\left(\eta c e B^\prime\right)$; see, {\it e.g.}, Ref.~\cite{Baerwald:2013pu}). However, high efficiencies are required to reach the highest cosmic-ray energies, and we have found \cite{Baerwald:2013pu} that in this case ($\eta \approx 1 $) the optically thin region vanishes: therefore, we conclude that the basic neutron model usually assumed in the literature is plausible only at unrealistically low acceleration efficiencies, which constitutes a further reason to disfavor it. 

Finally, as detailed in Ref.~\cite{Baerwald:2014zga}, we have considered a population of GRBs that evolves in redshift according to the star formation rate, and performed a self-consistent analysis that makes joint use of the flux at Earth of the UHECRs, of the prompt neutrinos created at the sources, and of the cosmogenic neutrinos created in the $p\gamma$ interactions between the UHECR protons and the photons of the cosmic microwave background and of the infrared/optical background. Unlike the previous analyses, we now normalize the UHECR flux to the spectrum measured by HiRes \cite{Abbasi:2007sv}; by doing this, the prompt and cosmogenic neutrino spectra are automatically normalized. The propagation of protons from the sources to Earth is performed via a numerical Boltzmann solver which considers energy losses due to the adiabatic cosmological expansion, and to the photohadronic and pair-creation interactions with the cosmological photon backgrounds. 

\begin{figure}
  \includegraphics[width=.655\textwidth]{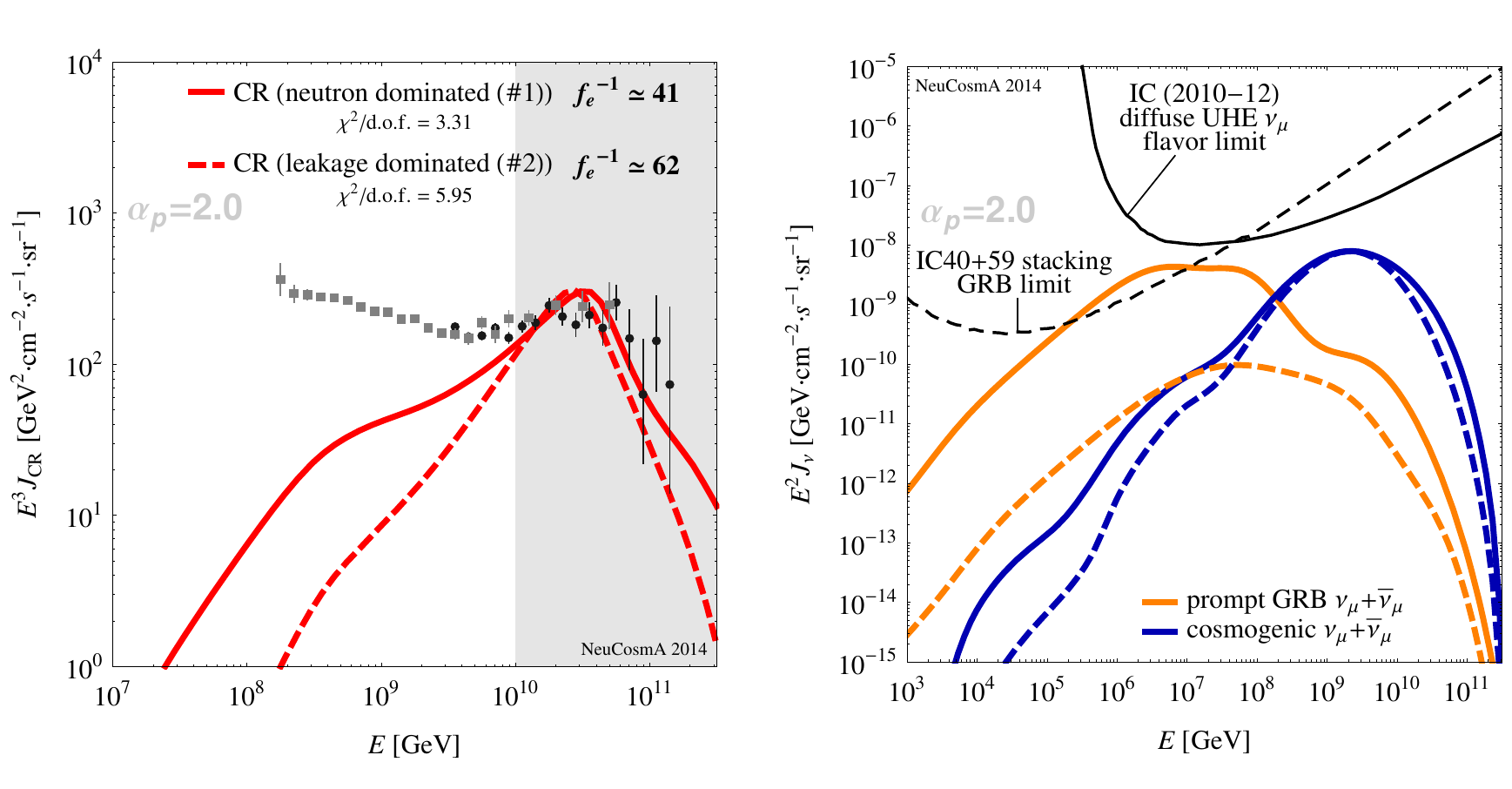}
  \caption{\label{fig:CRescapeModeldifferences}{\it Left panel:} UHECR proton flux normalized to HiRes data in the range $10^{10}$\---$10^{12}$ GeV, for a neutron escape-dominated model (\#1) and a direct proton escape-dominated model (\#2). {\it Right panel:} corresponding prompt neutrino (orange) and cosmogenic neutrino (blue) fluxes for the two models, compared to the IceCube stacking GRB upper limit and the diffuse UHE $\nu_\mu$ limit. Figure taken from Ref.~\cite{Baerwald:2014zga}.}
\end{figure}

In the left panel of Fig.~\ref{fig:CRescapeModeldifferences} we show two different predictions for the UHECR flux: one dominated by neutron escape (model \#1) and one dominated by direct proton escape, or leakage (model \#2). Both have been fitted to the HiRes data points in the energy interval between $10^{10}$ and $10^{12}$ GeV (gray shaded band). The fitting procedure fixes the necessary value of the baryonic loading $f_e^{-1}$. Note that, even though the fit for the neutron escape model is slightly better than for the direct escape model ({\it i.e.}, slightly lower reduced $\chi^2$), the right panel shows that the prompt neutrino flux associated to the neutron model exceeds the IceCube upper bound, while the one associated to direct escape lies safely below it. The reason is that in the latter case most of the highest-energy protons leave the source before interacting and creating neutrinos. The cosmogenic neutrino flux, on the other hand, is only mildly affected by the choice of UHECR model, since it depends only on the total number of neutrons plus protons that leave the source. By repeating this analysis for different points in the parameter space of GRB emission and propagation, we are able to find regions that are already excluded at present by the IceCube upper bounds, and others that will be excluded in ten to fifteen years of detector exposure, thus proving that a self-consistent analysis of cosmic rays and neutrinos is a powerful tool \cite{Baerwald:2014zga}.

\section{Summary and conclusions}

The current generation of neutrino telescopes is now reaching sensitivities that put the original predictions of the neutrino flux from GRBs to test. In fact, the latest analysis by IceCube disfavored the basic neutron model of UHECR and neutrino production. We have found, however, that a numerical recalculation involving more detailed particle physics results in a neutrino flux prediction that is still one order of magnitude safely below the IceCube bound \cite{Hummer:2011ms}. Furthermore, we have considered a generalized emission model that allows for UHECR injection via direct proton escape in addition to neutron escape, and with varying optical depth to neutron escape, which led us to find that, in order to reach the highest cosmic-ray energies, some amount of direct proton escape is necessary \cite{Baerwald:2013pu}. Finally, we performed a self-consistent joint analysis involving the UHECRs, prompt neutrinos, and cosmogenic neutrinos from GRBs \cite{Baerwald:2014zga}. By normalizing the UHECR flux at Earth to the spectrum measured by HiRes and by comparing the associated neutrino fluxes to the IceCube upper bounds, we conclude that it is already possible at present to find regions of the parameter space of GRB emission and propagation that are excluded. With more data to come in the following years in the cosmic-ray, neutrino, and gamma-ray channels, it seems possible that we could finally build a full picture of the connection between the UHE messengers of the Universe, and possibly to pinpoint their origin.






\begin{theacknowledgments}
 M.B. would like to thank the organizers of the workshop. This work was supported by the GRK1147 ``Theoretical Astrophysics and Particle Physics'', the FP7 Invisibles network (Marie Curie Actions, PITN-GA-2011-289442), the Helmholtz Alliance for Astroparticle Physics HAP, and DFG Grant WI 2639/4-1.
\end{theacknowledgments}



\bibliographystyle{aipproc}   


\begin{thebibliography}{15}
\expandafter\ifx\csname natexlab\endcsname\relax\def\natexlab#1{#1}\fi
\providecommand{\enquote}[1]{``#1''}
\expandafter\ifx\csname url\endcsname\relax
  \def\url#1{\texttt{#1}}\fi
\expandafter\ifx\csname urlprefix\endcsname\relax\def\urlprefix{URL }\fi
\providecommand{\eprint}[2][]{\url{#2}}

\bibitem[Abraham et~al.(2010)]{Abraham:2010yv}
J.~Abraham, et~al., \emph{Phys.Rev.Lett.} \textbf{104}, 091101 (2010),
  \eprint{1002.0699}.

\bibitem[Waxman(1995)]{Waxman:1995dg}
E.~Waxman, \emph{Astrophys.J.} \textbf{452}, L1--L4 (1995),
  \eprint{astro-ph/9508037}.

\bibitem[Waxman and Bahcall(1999)]{Waxman:1998yy}
E.~Waxman, and J.~N. Bahcall, \emph{Phys.Rev.} \textbf{D59}, 023002 (1999),
  \eprint{hep-ph/9807282}.

\bibitem[Ahlers et~al.(2011)]{Ahlers:2011jj}
M.~Ahlers, M.~Gonzalez-Garcia, and F.~Halzen, \emph{Astropart.Phys.}
  \textbf{35}, 87--94 (2011), \eprint{1103.3421}.

\bibitem[Abbasi et~al.(2012)]{Abbasi:2012zw}
R.~Abbasi, et~al., \emph{Nature} \textbf{484}, 351--353 (2012),
  \eprint{1204.4219}.

\bibitem[Guetta et~al.(2004)]{Guetta:2003wi}
D.~Guetta, D.~Hooper, J.~Alvarez-Muniz, F.~Halzen, and E.~Reuveni,
  \emph{Astropart.Phys.} \textbf{20}, 429--455 (2004),
  \eprint{astro-ph/0302524}.

\bibitem[Hummer et~al.(2012)]{Hummer:2011ms}
S.~Hummer, P.~Baerwald, and W.~Winter, \emph{Phys.Rev.Lett.} \textbf{108},
  231101 (2012), \eprint{1112.1076}.

\bibitem[Hummer et~al.(2010)]{Hummer:2010vx}
S.~Hummer, M.~Ruger, F.~Spanier, and W.~Winter, \emph{Astrophys.J.}
  \textbf{721}, 630--652 (2010), \eprint{1002.1310}.

\bibitem[Baerwald et~al.(2011)]{Baerwald:2010fk}
P.~Baerwald, S.~Hummer, and W.~Winter, \emph{Phys.Rev.} \textbf{D83}, 067303
  (2011), \eprint{1009.4010}.

\bibitem[Baerwald et~al.(2012)]{Baerwald:2011ee}
P.~Baerwald, S.~Hummer, and W.~Winter, \emph{Astropart.Phys.} \textbf{35},
  508--529 (2012), \eprint{1107.5583}.

\bibitem[He et~al.(2012)]{He:2012tq}
H.-N. He, R.-Y. Liu, X.-Y. Wang, S.~Nagataki, K.~Murase, et~al.,
  \emph{Astrophys.J.} \textbf{752}, 29 (2012), \eprint{1204.0857}.

\bibitem[Adri\'an-Mart\'inez et~al.(2013)]{Adrian-Martinez:2013sga}
S.~Adri\'an-Mart\'inez, A.~Albert, I.~A. Samarai, M.~Andr\'e, M.~Anghinolfi,
  et~al., \emph{A\&A 559,} \textbf{A9} (2013), \eprint{1307.0304}.

\bibitem[Baerwald et~al.(2013)]{Baerwald:2013pu}
P.~Baerwald, M.~Bustamante, and W.~Winter, \emph{Astrophys.J.} \textbf{768},
  186 (2013), \eprint{1301.6163}.

\bibitem[Baerwald et~al.(2014)]{Baerwald:2014zga}
P.~Baerwald, M.~Bustamante, and W.~Winter  (2014), \eprint{1401.1820}.

\bibitem[Abbasi et~al.(2008)]{Abbasi:2007sv}
R.~Abbasi, et~al., \emph{Phys.Rev.Lett.} \textbf{100}, 101101 (2008),
  \eprint{astro-ph/0703099}.

\end{thebibliography}



\end{document}